\documentclass[preprint,showpacs,preprintnumbers,amsmath,amssymb]{revtex4}
\usepackage{graphicx}
\usepackage{dcolumn}
\usepackage{bm}

\begin{document}


\title{A possible source of spin-polarized electrons:\\The inert graphene/Ni(111) system}

\author{Yu. S. Dedkov\footnote{Corresponding author. Electronic address: dedkov@physik.phy.tu-dresden.de} and C. Laubschat}
\affiliation{Institut f\"ur Festk\"orperphysik, Technische Universit\"at Dresden, 01062 Dresden, Germany}

\date{\today}

\begin{abstract}

We report on an investigation of spin-polarized secondary electron emission from the chemically inert system: graphene/Ni(111). An ordered passivated graphene layer (monolayer of graphite, MG) was formed on Ni(111) surface via cracking of propylene gas. The spin-polarization of the secondary electrons obtained from this system upon photoemission is only slightly lower than the one from the clean Ni surface, but does not change upon large oxygen exposure. These results suggest to use such passivated Ni(111) surface as a source of spin-polarized electrons stable against adsorption of reactive gases. 

\end{abstract}


\maketitle

Recently, graphene (monolayer of graphite, MG) has attracted renewed interest due to the possible application of this material in carbon-based nanoelectronics. It was shown theoretically~\cite{Lukyanchuk:2004} as well as experimentally~\cite{Novoselov:2005,Zhang:2005} that electrons move through graphene sheets as if they have no rest masses. In the same experimental work an unusual form of the quantum Hall effect was also observed. The peculiar band structure of graphene, a single $sp^2$ bonded carbon layer, possesses conical electron and hole pockets which meet only at the $K$-points of the Brillouin zone in momentum space~\cite{Novoselov:2005,Zhou:2006}. Due to the linear dependence of energy on momentum, the carriers behave as effectively massless, relativistic Dirac-fermions with an effective speed of light of $c_{eff}=10^6$\,m/s as described by Dirac's equation. From a technological point of few the observation of large carrier mobilities of up to 60000\,cm$^2$V$^{-1}$s$^{-1}$ at 4\,K and 15000\,cm$^2$V$^{-1}$s$^{-1}$ at 300\,K together with an bipolar field effect~\cite{Novoselov:2004} is most intriguing. For semiconductor-electronic applications graphite layers can be grown on SiC(0001) surfaces in high structural quality. In recent work similar intriguing properties in ultra-thin graphite layers grown on SiC was found and it was suggested that this may open a route toward new, graphene-based electronics~\cite{Berger:2006}. Weakly-bonded with substrate graphene layers can also be prepared on top of a Ni(111) substrate followed by intercalation of thin layers of noble metals (Cu, Ag, Au) underneath graphene layer~\cite{Dedkov:2001,Farias:1999,Shikin:2000}.

On the other hand stable non-reactive graphene layers on top of ferromagnetic materials (Ni)~\cite{Dedkov:2001,Farias:1999,Shikin:2000} may be used as sources of spin-polarized electrons. Electron sources are used in all domains ranging from technical devices of daily life like cathode-ray tubes to large-scale scientific experiments like electron accelerators. While the energy distribution and the average kinetic energy of the electrons can easily be controlled by fine tuning of the electron emission parameters (like, e.g., bias potential and temperature of the source), the control over the spin polarization of the electron beam is difficult. The latter is of great interest for particle physics experiments and for studies of magnetic systems in condensed matter physics, including the burgeoning field of spintronics.

In the present work we report on spin-polarized electron emission from graphene/Ni(111) system before and after exposure to oxygen. It is shown that the spin-polarization of secondary electrons at zero kinetic energy from graphene/Ni(111) system is reduced by about $1/3$ with respect to one from the clean Ni(111) surface, but contrasting to the latter it remains practically unaffected upon oxygen exposure. These experimental observations opens technical perspectives for application of graphite layers in spintronic devices. 

Investigations of the graphene/Ni(111) system were performed in the experimental setup for spin-resolved electron spectroscopy consisting of two chambers: preparation and analysis. As a substrate the W(110) single crystal was used. Prior to preparation of the studied system the well established cleaning procedure of the W-substrate was applied: several cycles of oxygen treatment with subsequent flashes at 2300$^\circ$\,C. A well ordered Ni(111) surface was prepared by thermal deposition of Ni films with a thickness of about 100\,\AA\ on to a clean W(110) substrate and subsequent annealing at 300$^\circ$\,C. The corresponding LEED pattern is shown in the left upper conner of Fig.\,1. An ordered graphene overlayer was prepared via cracking of propylene gas (C$_3$H$_6$) according to the recipe described in Ref.~\cite{Dedkov:2001}. The LEED spots of the graphene/Ni(111) system reveal a well-ordered $p(1\times1)$-overstructure as expected from the small lattice mismatch of only 1.3\% (Fig.\,1, upper panel, center). Spin- and angle-resolved photoemission spectra were recorded at 1486.6\,eV (Al\,K$\alpha$, XPS) and 40.8\,eV (He\,II$\alpha$, UPS) photon energy, respectively, using a hemispherical energy analyzer SPECS PHOIBOS 150 combined with a 25\,kV mini-Mott detector for spin analysis~\cite{specs}. The energy resolution of the analyzer was set to 100 and 500\,meV for UPS and XPS, respectively. The spin polarization of secondary electrons in XPS spectra was analyzed with an energy resolution of 200\,meV. Spin-resolved measurements were performed in magnetic remanence after having applied a magnetic field pulse of about 1\,kOe along the in-plane $<1\bar{1}0>$ easy axis of the Ni(111) film. The experimental setup asymmetry was accounted for in the standard way by measuring spin-resolved spectra for two opposite directions of applied magnetic field~\cite{Kessler:1985,Johnson:1992}. 

The electronic structure of the graphene/Ni(111) system was studied in detail by means of angle-resolved photoemission in earlier work~\cite{Dedkov:2001,Nagashima:1994}. Here we show only a few angle-resolved valence-band photoemission spectra of the system under study (Fig.\,1, lower panel). The spectra were taken with He\,II$\alpha$ radiation along the $\overline{\Gamma}-\overline{M}$ direction of the surface Brillouin zone (Fig.\,1, right upper conner) and are in good agreement with previous data~\cite{Dedkov:2001,Nagashima:1994}. In the same figure the photoemission spectrum of a pure graphite single crystal measured in normal emission geometry is shown by a shaded area. From a comparison of the photoemission spectra of graphene/Ni(111) and pure graphite one may conclude that the difference in binding energy of the $\pi$-states amounts to about 2.3\,eV which is close to the value observed earlier~\cite{Dedkov:2001,Nagashima:1994} and in good agreement with the theoretical prediction of 2.35\,eV~\cite{Bertoni:2004}. This shift reflects the effect of hybridization of the graphene $\pi$ bands with the Ni $3d$ bands and, secondary, with the Ni $4s$ and $4p$ states. These results indicate high quality of the graphene monolayer on top of the Ni(111) surface.

The inert properties of the graphene/Ni(111) system were tested by exposure to oxygen for 30\,minutes at a partial O$_2$-pressure of $5\times10^{-6}$\,mbar and room temperature. The results are compiled in Fig.\,2 where a series of Ni\,$2p$ XPS spectra are shown, taken from the pure Ni(111) surface (spectrum 1), from freshly prepared graphene/Ni(111) (spectrum 2), and after exposure of graphene/Ni(111) and Ni(111) to oxygen (spectra 3 and 4, respectively). The inset of Fig.\,2 shows O\,$1s$ XPS spectra obtained from the last two systems. In all spectra the Ni $2p$ emission consists of a spin-orbit doublet ($2p_{3/2,1/2}$) and a well-known satellite structure. The main line is ascribed to a completely screened final state $(c^{-1}3d^{10}4s^{1})$ and the satellite to a two-hole bound state $(c^{-1}3d^{9}4s^{2})$, where $c^{-1}$ stands for the [Ar] core with a $2p$ hole. For the pure Ni(111) film the satellite appears with respect to the main lines at 6\,eV higher binding energy, whereas this shift is increased approximately by 0.9\,eV for the graphene/Ni(111) system. This effect reflects the altered chemical environment at the interface and is not subject of the present discussion. From a comparison of spectra 2 and 3 it becomes clear that exposure to oxygen does not affect the spectral shape of the graphene/Ni(111) system. It shows that the graphene overlayer prevent obviously the interaction of oxygen with the underlying Ni substrate which would be reflected by strong modifications of satellite structure of the Ni $2p$ spectra (compare spectrum 4). The intensity of the O $1s$ photoemission signal of the graphene/Ni(111) system after oxygen exposure is very weak as compared to the one of the pure Ni surface upon the same treatment (compare inset of Fig.\,2). From this observation one may conclude that the sticking coefficient of oxygen on the graphene overlayer is extremely low and the overlayer is almost free of defects that allow oxygen atoms access to the Ni substrate. From these facts one may suggest that a Ni(111) surface passivated by graphene may serve as an extremely inert source of spin-polarized electrons.

This idea was tested experimentally by measuring the spin polarization, $P$, of the secondary electrons of the systems under study. Here $P$ is defined as normalized difference of the numbers of the emitted spin-up and spin-down electrons. The spin-polarization of the low kinetic energy tail of the photoelectron spectrum of the ferromagnetic Ni(111) film (see Fig.\,3, open circles) is in good agreement with previously published results~\cite{Hopster:1983,Landolt:1986,Kamper:1989,Allenspach:2000}. The spin polarization of secondary electrons at kinetic energies of about 10\,eV is equal to the one of the valence band electrons. A strong enhancement of the spin polarization appears at kinetic energies below $\approx3$\,eV. This enhancement was interpreted as due to exchange scattering between the hot electrons and the $3d$ valence electrons~\cite{Glazer:1984,Penn:1985}, the leading term being the ``spin-flip" processes of the Stoner excitations. Monte-Carlo simulations~\cite{Glazer:1984} showed that the strongly enhanced polarization near zero kinetic energy still scales with magnetization of the sample, $M$. The situation is very favorable for applications as surface magnetometer or source of spin-polarized electrons: the maximum of polarization coincides with the maximum of the intensity, $I$, and the intensity of true secondary electrons generally is very large, orders of magnitude larger than the intensities of elastically scattered electrons or photoelectrons.

Figure\,3 shows the spin-polarization of the secondary electrons as measured for the pure Ni(111) surface (open circles), graphene/Ni(111) (filled circles), O$_2$+graphene/Ni(111) (half-filled squares), and O$_2$+Ni(111) (open triangles). As compared to the clean Ni(111) surface, the spin polarization of secondary electrons emitted from the graphene/Ni(111) system is reduced from $\approx17$\% to $\approx12$\%. The reduction of $P$ due to the graphene overlayer can be explained by the strong hybridization between Ni $3d$ and graphene $\pi$ states that is supported by theory~\cite{Souzu:1995,Bertoni:2004} and previous angle-resolved photoemission experiments~\cite{Dedkov:2001,Nagashima:1994}. Since the spin-polarization of secondary electrons is rougly proportional to the magnetic moment at the surface of a ferromagnetic material~\cite{Hopster:1983,Landolt:1986,Kamper:1989,Allenspach:2000}, one can estimate, that in the graphene/Ni(111) system the magnetic moment of the Ni atoms at the interface amounts to about 0.52\,$\mu_B$ as compared to 0.72\,$\mu_B$ for the pure Ni(111) surface~\cite{Bertoni:2004}. This reduction is consistent with values obtained in a recent theoretical work~\cite{Bertoni:2004}. Additional reduction of the magnetic moment can be explained by spin-flip scattering of electrons in the polarized graphene layer or/and scattering by structural defects~\cite{Rutter:2007}.

The interesting observation is that exposure to oxygen at high partial pressure does not affect the shape of the spin-polarization curve. Particularly, the spin polarization of electrons at the vacuum level (zero kinetic energy) remains almost the same (the slight reduction by $~2$\% is within error bar). This is in strong contrast to the behavior of the clean Ni(111) surface where exposure to oxygen leads to a complete quenching of the spin polarization (compare Fig.\,3).

\textit{In conclusion}, we studied emission of spin-polarized secondary electrons from the graphene/Ni(111) system and found that exposure to large amounts of oxygen does not affect the magnitude of spin polarization. The phenomenon is explained by a passivation of the Ni surface by a closed graphene overlayer that inhibits oxygen adsorption and direct contact of Ni atoms with the reactive gas. We suggest that such an inert system may be used as chemically inert source of spin-polarized electrons.

This work was funded by the Deutsche Forschungsgemeinschaft, SFB 463 TP B4.

\newpage

\textbf{Figure captions:}
\newline
\newline
\textbf{Fig.\,1.}\,\,Upper panel, from left to right: LEED patterns from pure Ni(111) and graphene/Ni(111) surfaces, and corresponding surface Brillouin zone of the system with main symmetric points. Lower panel: Photoelectron spectra of the graphene/Ni(111) system taken with $h\nu=40.8$\,eV photon energy along $\overline{\Gamma}-\overline{M}$ direction of the surface Brillouin zone are presented for several emission angles (marked on the left-hand side of each spectra); for comparison a spectrum of pure graphite taken in normal emission geometry is shown by the shaded area. 
\newline
\newline
\textbf{Fig.\,2.}\,\,Ni 2\textit{p} core-level spectra of Ni(111), graphene/Ni(111), and systems obtained after exposure of the respective surfaces to oxygen. The inset shows O 1\textit{s} XPS spectra obtained after exposure of Ni(111) and graphene/Ni(111) surfaces to large amounts of oxygen, respectively.  
\newline
\newline
\textbf{Fig.\,3.}\,\,Spin-polarization of secondary electrons from Ni(111) (open circles), graphene/Ni(111) (filled circles), O$_2$+graphene/Ni(111) (half-filled squares), and O$_2$+Ni(111) (open triangles) systems after excitation with Al\,K$\alpha$ radiation. Solid lines are shown as guides to the eye.    

\clearpage
\begin{figure}[t]\center
\includegraphics[scale=0.7]{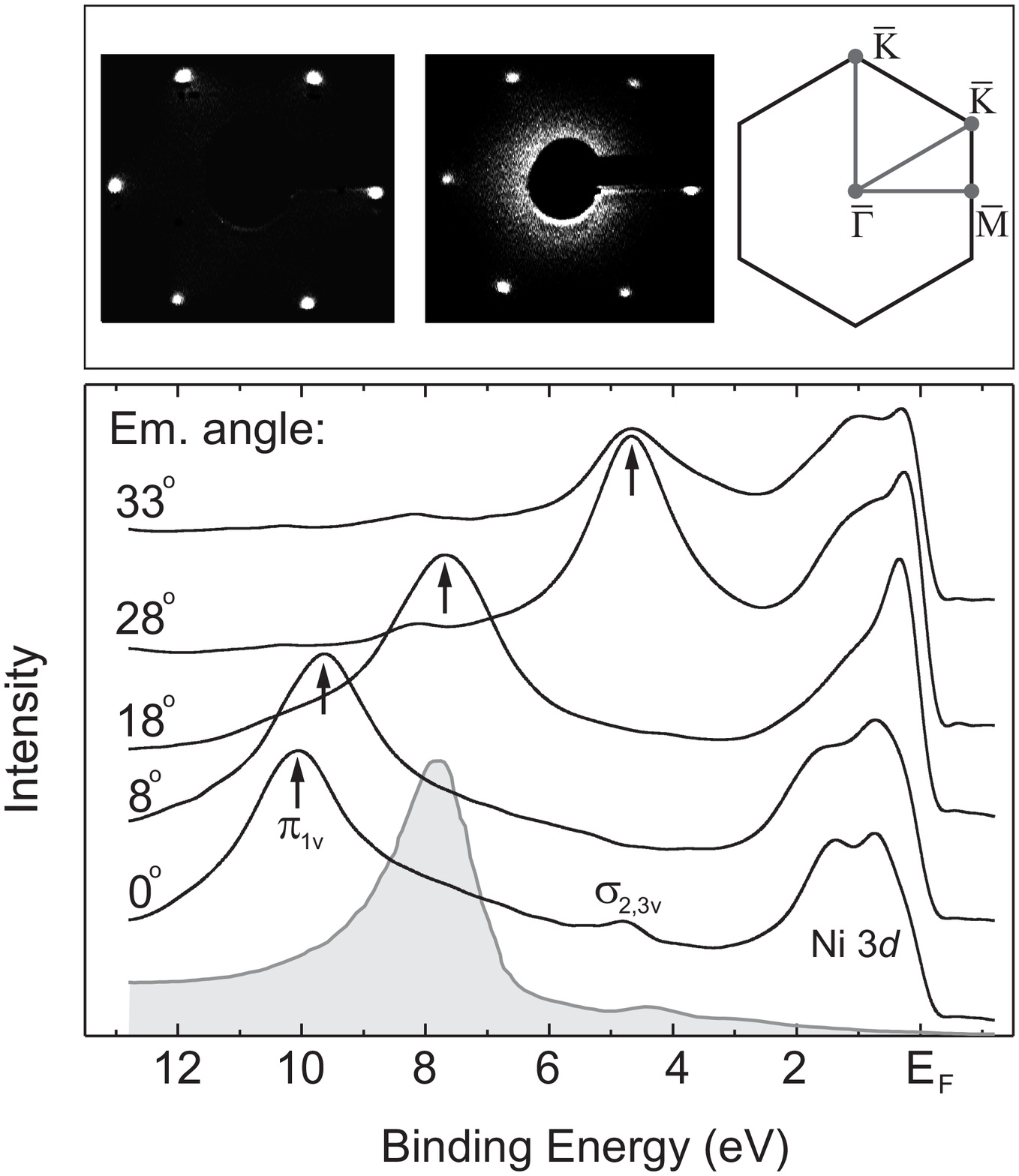}\\
\vspace{1cm}
\large \textbf{Fig.\,1, Yu. S. Dedkov \textit{et al.}, Appl. Phys. Lett.}
\end{figure}

\clearpage
\begin{figure}[t]\center
\includegraphics[scale=0.7]{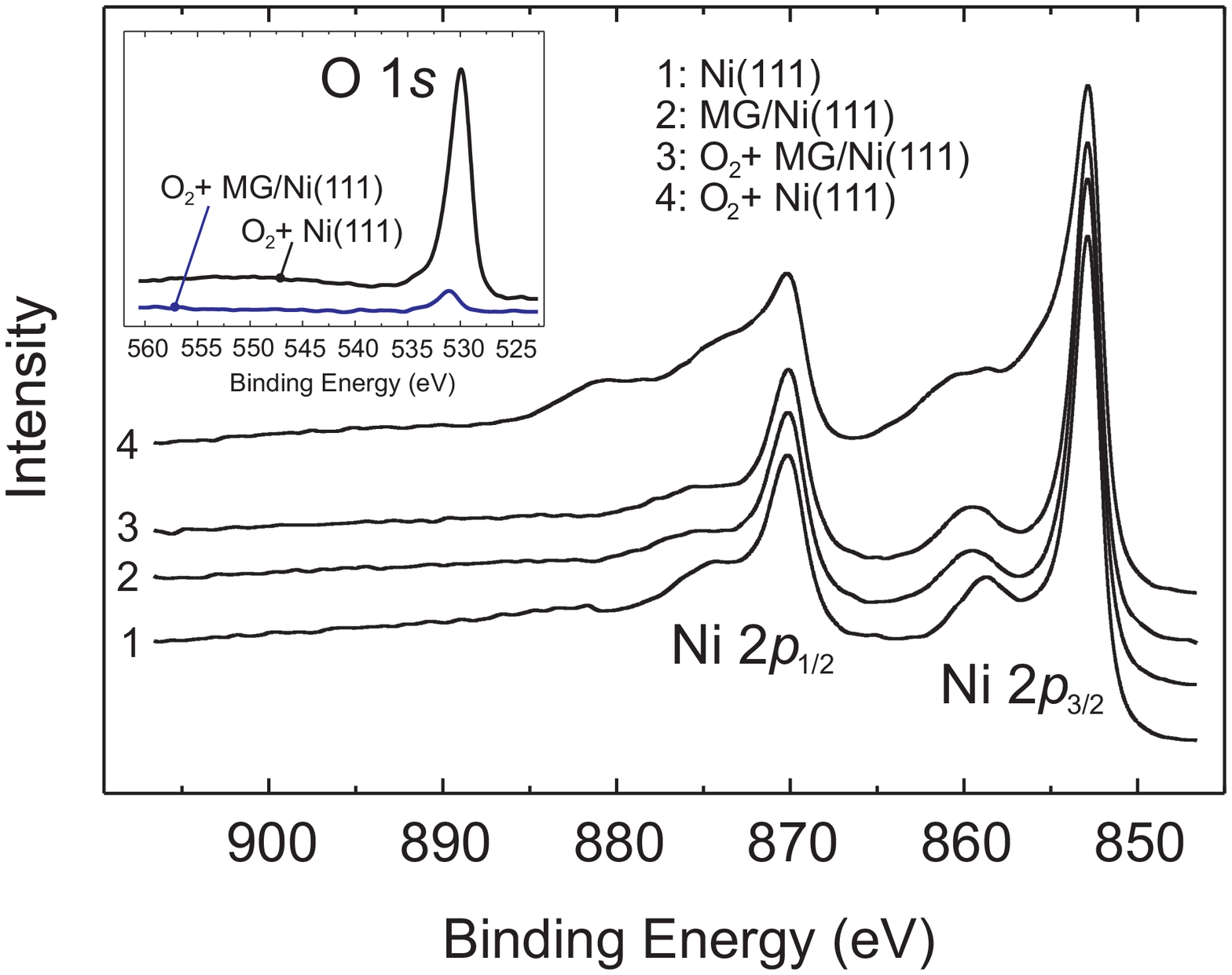}\\
\vspace{1cm}
\large \textbf{Fig.\,2, Yu. S. Dedkov \textit{et al.}, Appl. Phys. Lett.}
\end{figure}

\clearpage
\begin{figure}[t]\center
\includegraphics[scale=0.7]{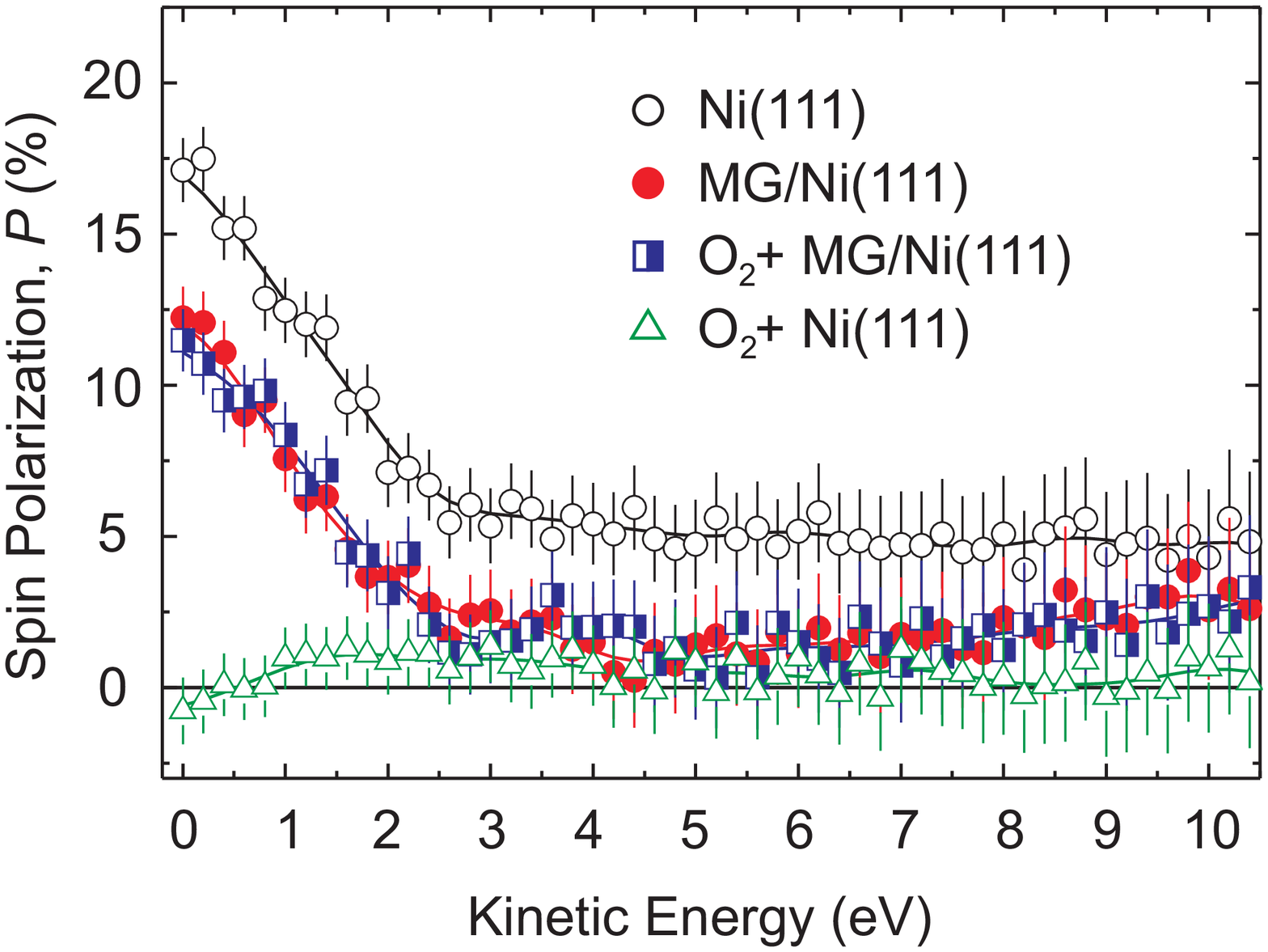}\\
\vspace{1cm}
\large \textbf{Fig.\,3, Yu. S. Dedkov \textit{et al.}, Appl. Phys. Lett.}
\end{figure}


\clearpage

\begin{thebibliography}{10}

\bibitem{Lukyanchuk:2004} I.A. Luk'yanchuk and Y. Kopelevich, Phys. Rev. Lett. \textbf{93}, 166402 (2004).

\bibitem{Novoselov:2005} K. S. Novoselov, A. K. Geim, S. V. Morozov, D. Jiang, M. I. Katsnelson, I. V. Grigorieva, S. V. Dubonos, and A. A. Firsov, Nature \textbf{438}, 197 (2005).

\bibitem{Zhang:2005} Y. Zhang, Y.-W. Tan, H. L. Stormer, and Philip Kim, Nature \textbf{438}, 201 (2005).

\bibitem{Zhou:2006} S.Y. Zhou, G.-H. Gweon, and A. Lanzara, Ann. Phys. \textbf{321}, 1730 (2006).

\bibitem{Novoselov:2004} K. S. Novoselov, A. K. Geim, S. V. Morozov, D. Jiang, Y. Zhang, S. V. Dubonos, I. V. Grigorieva, and A. A. Firsov, Science \textbf{306}, 666 (2004).

\bibitem{Berger:2006} C. Berger, Z. Song, X. Li, X. Wu, N. Brown, C. Naud, D. Mayou, T. Li, J. Hass, A. N. Marchenkov, E. H. Conrad, P. N. First, and W. A. de Heer, Science \textbf{312}, 1191 (2006).

\bibitem{Dedkov:2001} Yu. S. Dedkov, A. M. Shikin, V. K. Adamchuk, S. L. Molodtsov, C. Laubschat, A. Bauer, and G. Kaindl, Phys. Rev. B \textbf{64}, 035405 (2001).

\bibitem{Farias:1999} D. Farias, A. M. Shikin, K.-H. Rieder, and Yu. S. Dedkov. J. Phys.: Condens. Matter \textbf{11}, 8453 (1999).

\bibitem{Shikin:2000} A. M. Shikin, G. V. Prudnikova, V. K. Adamchuk, F. Moresco, and K.-H. Rieder, Phys. Rev. B \textbf{62}, 13202 (2000).

\bibitem{specs} http://www.specs.de.

\bibitem{Kessler:1985} J. Kessler, \textit{Polarized Electrons}, 2nd ed. (Springer-Verlag, Berlin, 1985).

\bibitem{Johnson:1992} P. D. Johnson \textit{et al.}, Rev. Sci. Instrum. \textbf{63}, 1902 (1992).

\bibitem{Nagashima:1994} A. Nagashima, N. Tejima, and C. Oshima, Phys. Rev. B \textbf{50}, 17487 (1994).

\bibitem{Bertoni:2004} G. Bertoni, L. Calmels, A. Altibelli, and V. Serin, Phys. Rev. B \textbf{71}, 075402 (2004).

\bibitem{Hopster:1983} H. Hopster, R. Raue, E. Kisker, G. G\"untherodt, M. Campagna, Phys. Rev. Lett. \textbf{50}, 71 (1983).

\bibitem{Landolt:1986} M. Landolt, Appl. Phys. A \textbf{41}, 83 (1986).

\bibitem{Kamper:1989} K.-P. K\"amper, Ph.D. dissertation, RWTH Aachen (1989).

\bibitem{Allenspach:2000} R. Allenspach, IBM J. Res. Dev. \textbf{44}, 553 (2000).

\bibitem{Glazer:1984} J. Glazer, E. Tosatfi, Solid State Commun. \textbf{52}, 905 (1984).

\bibitem{Penn:1985} D.R. Penn, S.P. Apell, S.M. Girwin, Phys. Rev. Lett. \textbf{55}, 518 (1985).

\bibitem{Souzu:1995} Y. Souzu and M. Tsukada, Surf. Sci. \textbf{326}, 42 (1995).

\bibitem{Rutter:2007} G. M. Rutter, J. N. Crain, N. P. Guisinger, T. Li, P. N. First, J. A. Stroscio, Science \textbf{317}, 219 (2007).

\end{thebibliography}
\end{document}